%% file: extver.tex
\documentclass[10pt,conference]{IEEEtran}
\IEEEoverridecommandlockouts
\usepackage{amsmath,amssymb,amsfonts,amsthm}
\usepackage{bbm}
\usepackage{cite}
\usepackage{graphicx}
\graphicspath{ {./figures/} }
\usepackage[linesnumbered,ruled,vlined]{algorithm2e}

\makeatletter
\def\input@path{{./figures/}}
\makeatother

\theoremstyle{plain}
\newtheorem{proposition}{Proposition}
\newtheorem{lemma}{Lemma}
\DeclareMathOperator*{\E}{\mathbb{E}}

\def\BibTeX{{\rm B\kern-.05em{\sc i\kern-.025em b}\kern-.08em
    T\kern-.1667em\lower.7ex\hbox{E}\kern-.125emX}}

\begin{document}

\title{Resource Efficiency vs Performance Isolation Tradeoff in Network Slicing
\thanks{.}
}

\author{\IEEEauthorblockN{Panagiotis Nikolaidis and Asim Zoulkarni and John Baras\\}
\IEEEauthorblockA{Department of Electrical \& Computer Engineering and the Institute for Systems Research\\
University of Maryland, College Park, MD 20742, USA\\
Email: \{nikolaid, asimz, baras\}@umd.edu}}

\maketitle

\begin{abstract}
We consider the tradeoff between resource efficiency and performance isolation that emerges when multiplexing the resource demands of Network Slices (NSs). On the one hand, multiplexing allows the use of idle resources, which increases resource efficiency. On the other hand, the performance of each NS becomes susceptible to traffic surges in other NSs, which degrades performance isolation. The analysis of this tradeoff enables network operators to determine the effect of performance isolation on the operating cost of each NS.

To study the tradeoff, we solve an optimization problem where we find the multiplexing policy that requires the least provisioned resources to honor the Service Level Agreements (SLAs) of all NSs. The SLA of each NS $i$ states that its resource demand should be met for $P^H_i$ fraction of time, and for $P^L_i \leq P^H_i$ fraction of time, it should be met regardless of the demands of other NSs.

For resource demands that follow ergodic Markov chains, we show that the well-known Max-Weight scheduler is an optimal multiplexing policy. Since the Max-Weight scheduler does not require any knowledge of the statistics of the resource demands, we also propose its use in non-markovian settings. For resource demands obtained in the LTE module of ns-3, we show that the Max-Weight scheduler reduces the provisioned bandwidth by $36.2\%$ when no performance isolation is required. Lastly, for these non-markovian resource demands, the Max-Weight scheduler maintains its optimality since it requires as much provisioned bandwidth as the best non-causal scheduler.
\end{abstract}

\begin{IEEEkeywords}
network slicing, resource efficiency, isolation, sharing, overbooking, tradeoff, Lyapunov, 5G, LTE, ns-3
\end{IEEEkeywords}

\section{Introduction}
\label{intro}
Future cellular networks need to serve a wide variety of applications. Examples include the traffic control of autonomous vehicles in intersections, the communications between ambulances and hospitals, and augmented reality. Such applications prioritize different types of Quality of Service (QoS) such as high reliability, low packet delays or high bitrates. 

These applications may be requested by different companies that may also require strict guarantees regarding the delivery of the desired QoS to their application. Therefore, each company can viewed as a "tenant" in the cellular network that forms a Service Level Agreement (SLA) with the Network Operator (NO) for the expected QoS delivered to their application. 

The above paradigm requires a cellular network that provides varying types of QoS with strict guarantees as stated in SLAs. This requirement has led to the emergence of network slicing, where the physical infrastructure of the cellular network is “sliced” to create multiple virtual networks called Network Slices (NSs). Each NS is an end-to-end network spanning over the Radio Access Network (RAN), the Transport Network (TN), and the Core Network (CN). 

To provide various types of QoS, the NO composes each NS using custom network functions that match the needs of its application. For instance, each NS may use its own Medium Access Control (MAC) scheduler in the RAN, routing policies in the TN and CN, and firewalls in the CN. The fast deployment of custom network functions has been enabled by virtualization techniques and software-defined networking.

To ensure that the SLA of each NS is honored, the NO needs to provision in advance enough resources so that the network functions of each NS can deliver the promised QoS. Based on the provisioned resources, the NO can then compute the cost of the SLA and provide a price quote to the tenant. Resource provisioning may involve traffic forecasting for each NS, however forecasting is out of the scope of this paper.

For instance, in the RAN, the NO may need to provision Physical Resource Blocks (PRBs) to the MAC scheduler of a NS to ensure high bitrates, or in the TN and CN, routing paths for to ensure high reliability.

Here, we are particularly interested in these provisioning problems. We wish to obtain the amount of resources needed so that the resource demand of a NS $i$ is met for a high fraction of time $P^H_i$. By resource demand, we refer to the amount of resources needed by a network function of a NS to provide the desired QoS given its current traffic state.

An obvious solution approach is to estimate the cdf of the resource demand of each NS and provision $P^H_i$-percentile resources for each of them. However, since the traffic state of a NS is time-varying, its resource demand also varies over time. Thus, provisioning $P^H_i$-percentile resources for each NS $i$ may be wasteful since some provisioned resources may remain unused for $P^H_i$ fraction of time.

This observation implies that allowing the unused provisioned resources of a NS $i$ to be used by some other NS $j$ may reduce the overall provisioned resources. Thus, it is beneficial to consider a scheduler that multiplexes the resource demands of NSs. Hence, the fulfillment of the SLAs in network slicing involves a joint scheduling and resource provisioning problem.

However, multiplexing may have negative effects on the performance of a NS $i$. For instance, suppose that a NS $i$ relies solely on the unused $P^H_i$-percentile provisions of the other NSs. Then, in case the other NSs experience unexpected traffic surges, e.g., due to DDoS attacks, the performance of NS $i$ will be severely affected. This is a highly undesirable outcome since tenants often require that the performance of their NS should remain unaffected by other NSs. This requirement is often referred to as performance isolation.

Here, we quantify performance isolation as the fraction of time $P^L_i$ that the resource demand of NS $i$ is guaranteed to be met regardless of the traffic state of other NSs. The higher $P^L_i$ is, the higher the degree of isolation is. A simple solution to provide $P^L_i$ degree of isolation to NS $i$ is to allocate $P^L_i$-percentile resources that are always available to NS $i$ if needed. 

Given the above, there is a tradeoff between resource efficiency and performance isolation. On one hand, scheduling resource demands reduces the provisioned resources and increases resource efficiency. On the other hand, heavily relying on the unused resources of other NSs degrades performance isolation as indicated in the previous DDoS attack example. The analysis of this tradeoff is the main topic of this paper.

We study this tradeoff by formulating a joint scheduling and resource provisioning optimization problem. The objective is to minimize the provisioned resources s.t. for each NS $i$, its resource demand is met for $P^H_i$ fraction of time, where for $P^L_i$ fraction of time, it is met regardless of the traffic conditions in other NSs. Adjusting $P^L_i$ in the above problem and computing the provisioned resources allows us to study the tradeoff.

We primarily focus on the instance of the problem involving the RAN part of a NS, where the network functions to be provisioned are the MAC schedulers of each NS and the resource demands are bandwidth demands measured in PRBs. Nonetheless, our approach can be applied to any network function whose resource provisioning decouples per node.

The main result of our paper is that the Max-Weight scheduler is an optimal multiplexing policy since it requires the least provisioned resources to satisfy the SLAs of all NSs, when their resource demands follow an ergodic Markov Chain (MC). To test its performance in non-markovian settings, we obtain resource demands from the LTE module of ns-3. The results show that the Max-Weight scheduler maintains its optimality and achieved considerable bandwidth savings.

\section{Related Work}
\label{relwork}
We are not the first to identify the resource efficiency vs performance isolation tradeoff. In \cite{imdea}, the authors studied the effect of overbooking strategies on resource allocation and service violations. Specifically, the authors considered the same $P^H$ and $P^L$ values for all NSs. Two scenarios where investigated; perfect sharing and network slicing. In perfect sharing, the BS simply sums the resource demands of all NSs and provisions $P^H$-percentile resources. In network slicing, performance isolation is considered by allocating dedicated $P^L$-percentile resources to each NS. The remaining resources needed to achieve $P^H-P^L$ fraction of time acceptance for each NS are computed based on past data. Although the above method provides significant insight, it can be used only when $P^H$ and $P^L$ are the same for all NSs, and also there is no insight on how to find the optimal multiplexing strategy.

A relevant problem was considered in \cite{dasilva}. There, the authors do not consider resource provisioning but assume that admission control ensures that the fixed available provisioned resources are less than the sum of the dedicated resources to each NS. The remaining resources that are not dedicated to any NS are considered auxiliary resources that are provided to each NS with probabilistic guarantees. The main topic of the paper is to provide efficient overbooking strategies for the auxiliary resources. The authors highlight the importance of forecasting methods and propose some interesting pricing models for this scenario. However, we believe that to properly analyze the tradeoff we are interested in, the joint consideration of overbooking and resource provisioning is needed.

The tradeoff was also considered to some extent in \cite{draig}. The authors study the effect of $P^H_i$ on the provisioned resources by solving at each timeslot $t$ a bandwidth minimization problem. However, multiplexing is not considered and the tradeoff cannot be fully studied. Also, solving a separate optimization problem each timeslot $t$ may lead to SLA violations over a long period of time if the resource demands are not iid.

Lastly, we mention \cite{isolation}. There, the authors describe the concept of isolation and provide means of implementing it for various network functions in the RAN, TN, and CN. It is stated that isolation can be considered in terms of performance, security, and dependability. In this paper, we are primarily concerned about performance isolation as the title suggests.

\section{System Architecture}
\label{sysa}
We consider $N$ NSs served by a single BS in a RAN. Each NS may have different type of QoS requirements. For instance, a NS may wish to upper bound the average packet delay of its users, whereas another NS may wish to provide constant bitrates to its users. For this reason, we also consider that NSs may use different MAC schedulers. 

Let vector $\mathbf{W}(t)=[W_1(t),W_2(t),...,W_N(t)]$ contain the bandwidth demands at timeslot $t$ of all $N$ NSs. By bandwidth demand $W_i(t)$, we refer to the number of PRBs that the MAC scheduler of NS $i$ needs in order to provide the desired QoS of NS $i$ throughout timeslot $t$. 

The determination of the bandwidth demand $W_i(t)$ given the current traffic state $s_i(t)$ of NS $i$ is not trivial, especially for complex QoS requirements. However, for simpler QoS metrics, it may be straightforward. For instance, suppose NS $i$ needs to deliver constant bitrates to each of its users. Then, the bandwidth demand $W_i(t)$ is computed using the Modulation and Coding Scheme (MCS) of each user at timeslot $t$. 

\begin{figure}
\centering
\includegraphics[width=\linewidth]{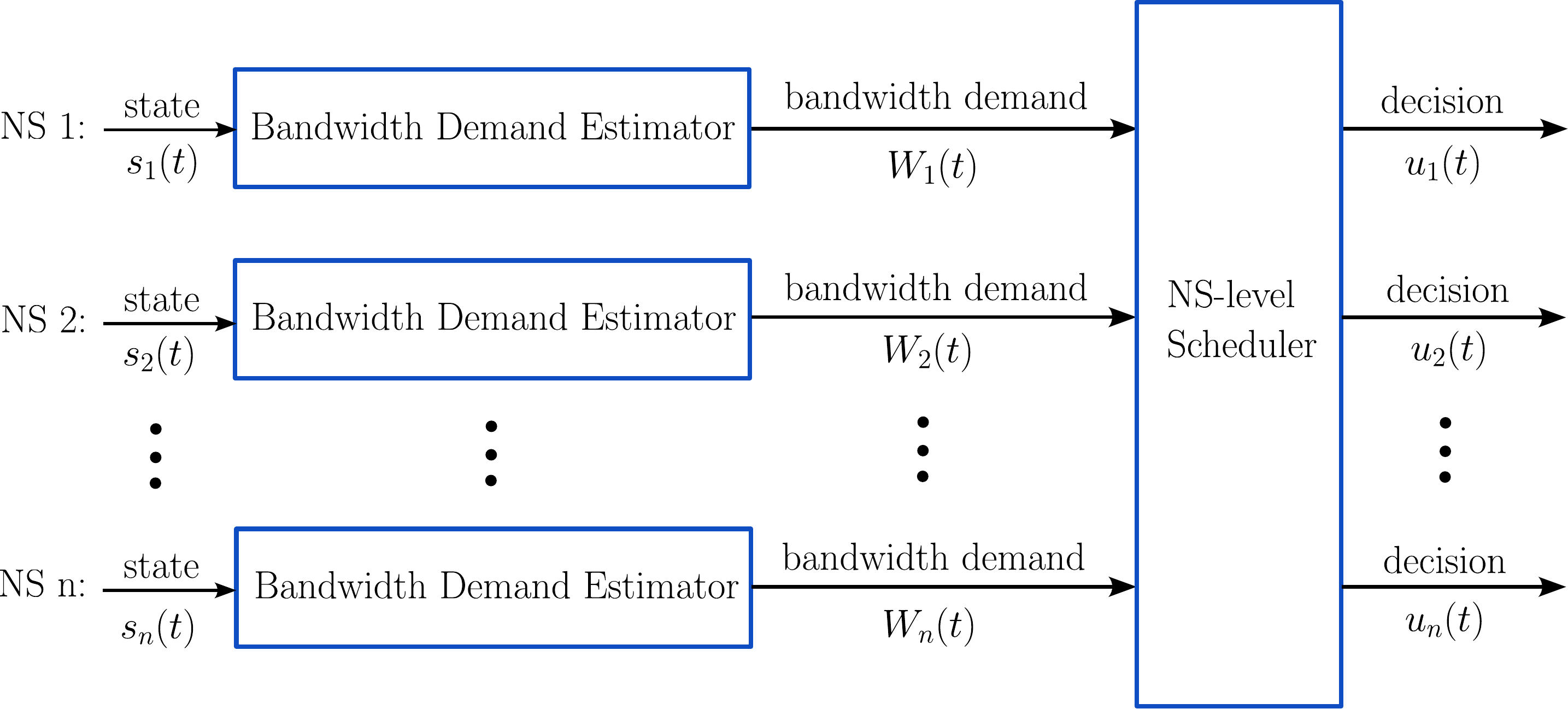}
\caption{In our system architecture, we consider two network functions. First, the Bandwidth Demand Estimator observes the traffic state $s_i(t)$ of a NS and estimates the number of PRBs $W_i(t)$ needed to meet its desired QoS. Second, the NS-level Scheduler receives all bandwidth demands $W_i(t)$ and decides which ones to satisfy given the limited bandwidth at the BS. The timeslot length depends on the timescales supported by these two network functions.}
\label{system}
\end{figure}

In any case, we consider that the bandwidth demand vector $\mathbf{W}(t)$ is estimated at each timeslot $t$ by a network function called bandwidth demand estimator whose operation is outside of the scope of this paper. Next, we consider a NS-level scheduler that at each timeslot $t$ decides whether to allocate the $W_i(t)$ PRBs to the MAC scheduler of NS $i$ given the overall demand vector $\mathbf{W}(t)$ and the limited bandwidth at the BS. Thus, the scheduler's output is a binary decision vector $\mathbf{u}(t)$. Figure \ref{system} depicts the system architecture.

% For the remainder of the paper, we focus on the NS-level scheduler. We wish to derive a scheduling policy that allows various degrees of performance isolation and requires minimum bandwidth at the BS to meet the SLAs of all NSs.

\section{Problem Formulation}
\label{probform}
The overall objective of our system is to satisfy a high percentage of the bandwidth demands $\mathbf{W}(t)$ over a long period of time. Considering that $u_i(t)=1$ if and only if the demand $W_i(t)$ is met, then we wish to satisfy the following constraint:
\begin{equation}
\liminf\limits_{T \to \infty}\frac{1}{T}\sum\limits_{t=1}^{T}u_i(t) \geq P_i^H, \: \forall i.
\label{SLA}
\end{equation}

Constraint (\ref{SLA}) states that NS $i$ needs to receive its desired QoS for a high percentage of time $P_i^H$. This type of constraints are known as availability constraints and are  widely used in real networks for bandwidth provisioning purposes, as in Google's software-defined network B4 \cite{after}.

Next, we consider that the scheduler provisions $W_i^r$ bandwidth at the BS for each NS $i$. Therefore, regardless of the traffic in the other NSs, the demand of NS $i$ is always met if it is less than $W_i^r$. Thus, bandwidth $W_i^r$ affects the degree of performance isolation that NS $i$ enjoys. Here, we consider the following performance isolation constraint:
\begin{equation}
\liminf\limits_{T \to \infty}\frac{1}{T}\sum\limits_{t=1}^{T}\mathbbm{1}_{W_i(t) \leq W_i^r} \geq P_i^L, \: \forall i, \mbox{ where } P_i^L < P_i^H.
\label{isolcon}
\end{equation}

Note that $P^L_i$ is the guaranteed percentage of time that NS $i$ receives its desired QoS regardless of unexpected traffic surges in other NSs. Thus, $P^L_i$ can be viewed as the degree of performance isolation that NS $i$ enjoys. The values of $P^H_i$ and $P^L_i$ are specified in the SLA between tenant $i$ and the NO. 

Next, note that since $u_i(t)=1$ if $W_i(t) \leq W_i^r$, constraint (\ref{SLA}) is equivalently reformulated as follows:
\begin{equation}
\liminf\limits_{T \to \infty}\frac{1}{T}\sum\limits_{t=1}^{T}\mathbbm{1}_{W_i(t) \leq W_i^r}+u_i(t)\mathbbm{1}_{W_i(t) > W_i^r}  \geq P_i^H, \: \forall i.
\label{SLA1}
\end{equation}

In case $W_i(t) \leq W_i^r$, we allow the scheduler to allocate the unused bandwidth provisions $W_i^r-W_i(t)$ of NS $i$ to some other NS $j$ that needs it, where $W_j(t) > W_j^r$. Thus, we allow the scheduler to multiplex the bandwidth demands of NSs.

Note that NS $i$ may reduce its $W_i^r$ provisions by relying on the unused provisions $\sum_{j \neq i}(W^r_j-W_j(t))$ of the other NSs. Thus, multiplexing increases resource efficiency since the overall provisioned bandwidth is reduced. On the other hand, if a NS relies heavily on the unused resources of other NSs, unexpected traffic surges in the other NSs will affect its performance, which degrades performance isolation.

\begin{figure}
\centering
\includegraphics[width=\linewidth]{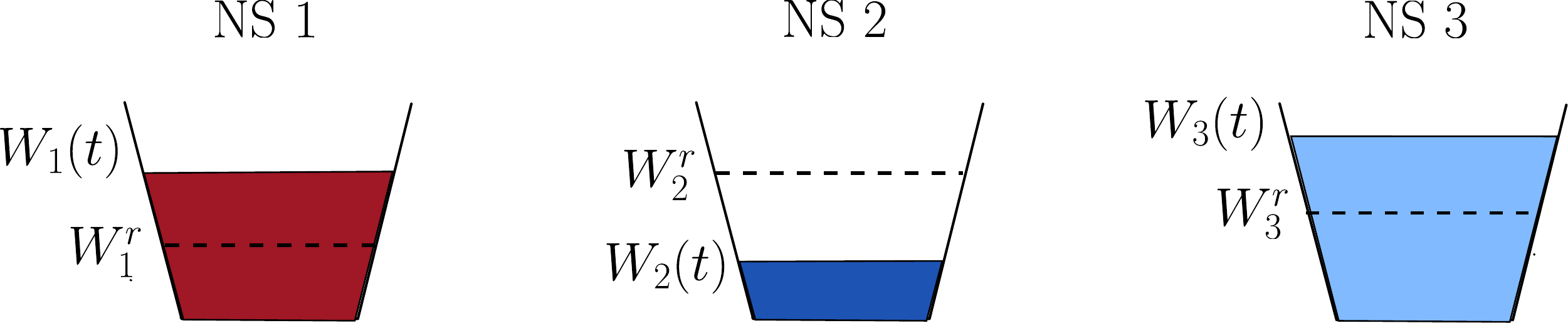}
\caption{Whenever $W_i(t) \leq W^r_i$, the demand of NS $i$ is met regardless of traffic in other NSs. In case $W_i(t) > W_i^r$, the scheduler may allocate the unused resources of the other NSs to NS $i$. Here, the unused resources of NS 2 do not suffice for the excess demands of both NS 1 and NS 3. Thus, the scheduler needs to decide which one to satisfy. Note that the larger $W_i^r$ is, the less frequently NS $i$ needs to rely on the unused resources of other NSs. Thus, increasing $W_i^r$ increases performance isolation. However, increasing $W_i^r$ also increases the provisioned bandwidth, hence the tradeoff.}
\label{multiplex}
\end{figure}

In the multiplexing paradigm, the scheduler decides at each time $t$ which excess bandwidth demands to satisfy given the available bandwidth. Thus, the scheduler is constrained by\footnote{We use the shorthand notation $[\cdot]^+$ for $\max\{\cdot,0\}$.}:
\begin{equation}
\mathbf{u}(t)^{\top}\left[\mathbf{W}(t)-\mathbf{W}^r\right]^+ \leq W^c + \mathbf{1}^\top \left[\mathbf{W^r}-\mathbf{W}(t)\right]^+, \; \forall t,
\label{optcriterion}
\end{equation}
where $W^c$ is an auxiliary bandwidth that is under the full control of the scheduler. Although the introduction of $W^c$ is not necessary, it helps with the analysis later on. The multiplexing paradigm is shown in Fig. \ref{multiplex}.

Next, note that $\mathbf{W(t)}$ may be a random vector given by some traffic forecasting model. Thus, to find the scheduler that needs the least provisioned bandwidth to meet each availability (\ref{SLA1}) and isolation (\ref{isolcon}) constraint, we need to solve: 
\begin{align}
& \underset{W^c, \mathbf{W^r}, \{\mathbf{u}(t)\}_{t \in \mathbb{N}}}{\text{minimize}}  W^c + \mathbf{1}^\top \mathbf{W^r}  \nonumber \\
& \text{s.t.:} \: \rm{Pr}\bigg{(}\liminf\limits_{T \to \infty}\frac{1}{T}\sum\limits_{t=1}^{T}\mathbbm{1}_{W_i(t) \leq W_i^r}+u_i(t)\mathbbm{1}_{W_i(t) > W_i^r}  \geq P_i^H \nonumber \\ 
& \phantom{\text{s.t.:} \:} \bigcap \: \liminf\limits_{T \to \infty}\frac{1}{T}\sum\limits_{t=1}^{T}\mathbbm{1}_{W_i(t) \leq W_i^r} \geq P_i^L \bigg{)} = 1, \: \forall i, \nonumber \\
& \phantom{\text{s.t.:} \:} \mathbf{u}(t)^{\top}\left[\mathbf{W}(t)-\mathbf{W}^r\right]^+ \leq W^c + \mathbf{1}^\top \left[\mathbf{W^r}-\mathbf{W}(t)\right]^+, \; \forall t, \nonumber \\
& \phantom{\text{s.t.:} \:} \mathbf{u}(t) \in \{0,1\}^N, \; \forall t. \label{schedprob}
\end{align}

The probabilistic constraint allows us to neglect extreme cases that occur w.p.0 that may result in over-provisioning in practice. Next, to simplify the first constraint, we assume:
\begin{equation}
\Pr\left(\liminf\limits_{T \to \infty}\frac{1}{T}\sum\limits_{t=1}^{T}\mathbbm{1}_{W_i(t) \leq W_i^r} = F_{W_i}(W_i^r)\right)=1, \: \forall W^r_i, \: \forall i.
\label{assumption}
\end{equation}

Assumption (\ref{assumption}) states that the time-average converges w.p.1 to a constant that depends on $W_i^r$. It implies that a single realization provides information for all time-averages over all sample paths. All ergodic processes $\mathbf{W}(t)$ satisfy (\ref{assumption}). Now, note that $F_{W_i}$ must be an increasing function and consider:
\begin{equation}
W_i^L \triangleq \min\{W_i^r \in \mathbb{R} : F_{W_i}(W_i^r) \geq P_i^L \}.
\label{wl}
\end{equation}

Bandwidth $W^L_i$ can be obtained through binary search by computing the time-average in (\ref{assumption}) through simulations due to the monotonicity of $F_{W_i}$. Next, note that the probability in the first constraint does not change if we intersect its events with the event in (\ref{assumption}). Thus, it can be seen that (\ref{schedprob}) is equivalent to:
\begin{align}
& \underset{W^c, \mathbf{W^r}, \{\mathbf{u}(t)\}_{t \in \mathbb{N}}}{\text{minimize}} W^c + \mathbf{1}^\top \mathbf{W^r}  \nonumber\\
& \text{s.t.:} \: \liminf\limits_{T \to \infty} \! \frac{1}{T} \!   \sum\limits_{t=1}^{T} \! u_i(t)\mathbbm{1}_{W_i(t) > W_i^r} \! \! \geq \! \! P_i^H \! \! - \! F_{W_i}(W_i^r),  \, w.p.1, \forall i, \nonumber\\
& \phantom{\text{s.t.:} \:} W_i^r \geq W_i^L,  \; \forall i, \nonumber \\
& \phantom{\text{s.t.:} \:} \mathbf{u}(t)^{\top}\left[\mathbf{W}(t)-\mathbf{W}^r\right]^+ \leq W^c + \mathbf{1}^\top \left[\mathbf{W^r}-\mathbf{W}(t)\right]^+, \; \forall t, \nonumber\\
&  \phantom{\text{s.t.:} \:} \mathbf{u}(t) \in \{0,1\}^N, \; \forall t. 
\label{schedprob1}
\end{align}

Problem (\ref{schedprob1}) is a joint scheduling and bandwidth provisioning problem with probabilistic time-average constraints. Its solution involves the determination of a scheduler  $\{\mathbf{u}(t)\}_{t \in \mathbb{N}}$ that satisfies the constraints with the least total bandwidth $W^c + \mathbf{1}^\top \mathbf{W^r}$ and the derivation of bandwidths $W^c$ and $\mathbf{W^r}$. 

Lastly, note that by adjusting the degree of performance isolation $P_i^L$, we affect $W_i^L$ in (\ref{wl}) which in turn affects the optimal value of (\ref{schedprob1}), i.e., the provisioned bandwidth. Thus, problem (\ref{schedprob1}) allows us to study the resource efficiency vs performance isolation tradeoff. In the next section, we transform (\ref{schedprob1}) to simpler forms until it becomes tractable.

\section{Equivalent Transformations}
\label{equivtrans}
From now on, we use the shorthand notation $\mathbf{u}$ for $\{\mathbf{u}(t)\}_{t \in \mathbb{N}}$. Due to the second constraint in (\ref{schedprob1}), all feasible solutions can be written as $(W^c, \mathbf{W^L} +\mathbf{e},\mathbf{u})$, where $\mathbf{e}$ is a vector with non-negative components. To simplify (\ref{schedprob1}), we observe that the existence of $W^c$ allows us to set $\mathbf{W^r}=\mathbf{W^L}$ and proceed without loss of optimality. To prove this, we first show the following proposition.
\begin{proposition}
Let $\mathbf{e} \in \mathcal{R}^N_+$. If $(W^c, \mathbf{W^L} +\mathbf{e},\mathbf{u})$ optimally solves (\ref{schedprob1}), then $(W^c + \mathbf{1}^{\top}\mathbf{e},\mathbf{W^L}, \mathbf{v})$ also optimally solves (\ref{schedprob1}), where $v_i(t)=1$ if $u_i(t)\mathbbm{1}_{W_i(t) > W_i^L + e_i}=1$ or $W_i^L \leq W_i(t) \leq W^L_i +e_i$, otherwise $v_i(t)=0$.

\noindent Proof: See Appendix \ref{appa}.
\label{simplify}
\end{proposition}
Proposition \ref{simplify} states that if the provisioned bandwidth $W^r_i$ is bigger than $W^L_i$ by some $e_i$, this $e_i$ can be instead added to the auxiliary bandwidth $W^c$ and still there exists a scheduler $\mathbf{v}$ that satisfies the constraints. Let $W_i^e(t) \triangleq W_i(t) - W_i^L$ denote the excess demand of NS $i$  when $W_i^r=W_i^L$, and let $P^M_i \triangleq F_{W_i}(W_i^L)$. We mention that if $F_{W_i}$ is continuous, then $P^M_i=P^L_i$ due to (\ref{wl}). By adding the constraint $\mathbf{W^r}=\mathbf{W^L}$ in (\ref{schedprob1}), we obtain the following problem:
\begin{align}
& \underset{W^c, \mathbf{u}}{\text{minimize}} \qquad W^c \nonumber\\
& \text{s.t.:} \: \liminf\limits_{T \to \infty}\frac{1}{T}\sum\limits_{t=1}^{T}u_i(t)\mathbbm{1}_{W_i^e(t) >0} \geq P_i^H - P_i^M,  \: w.p.1, \: \forall i, \nonumber\\
& \phantom{\text{s.t.:} \:} \mathbf{u}(t)^{\top}\left[\mathbf{W^e}(t)\right]^+ \leq W^c + \mathbf{1}^\top \left[-\mathbf{W^e}(t)\right]^+, \; \forall t, \nonumber\\
& \phantom{\text{s.t.:} \:} \mathbf{u}(t) \in \{0,1\}^N, \: \forall t.
\label{schedprob2}
\end{align}

\begin{proposition}
If $(W^c,\mathbf{W}^r,\mathbf{u})$ optimally solves (\ref{schedprob1}), then $(W^c +\mathbf{1}^\top\mathbf{e},\mathbf{v})$ optimally solves (\ref{schedprob2}), where $\mathbf{e} = \mathbf{W^r} - \mathbf{W^L}$ and $\mathbf{v}$ as defined in Proposition \ref{simplify}. If $(W^c,\mathbf{u})$ optimally solves (\ref{schedprob2}), then $(W^c,\mathbf{W}^L,\mathbf{u})$ optimally solves (\ref{schedprob1}).

\noindent Proof: See Appendix \ref{appb}.
\label{equiv1}
\end{proposition}
Thus, problems (\ref{schedprob1}) and (\ref{schedprob2}) are equivalent. However, problem (\ref{schedprob2}) has $N$ less optimization variables than (\ref{schedprob1}). To solve (\ref{schedprob2}), we need to find a scheduler $\mathbf{u}$ that satisfies the constraints using the smallest possible amount of bandwidth $W^c$ and compute this bandwidth $W^c$. Let $\mathcal{G}(W^c,\mathbf{P^H},\mathbf{P^M},\{\mathbf{W^e}(t)\}_{t \in \mathbb{N}})$ denote the set of schedulers $\mathbf{u}$ that satisfy the constraints of (\ref{schedprob2}) for a fixed value of $W^c$. We use the shorthand notation $\mathcal{G}_{W^c}$ for $\mathcal{G}(W^c,\mathbf{P^H},\mathbf{P^M},\{\mathbf{W^e}(t)\}_{t \in \mathbb{N}})$. Then, problem (\ref{schedprob2}) can be rewritten more compactly as:
\begin{equation}
\underset{W^c, \mathbf{u}}{\text{minimize}} \: \: W^c \: \: s.t. \: \: \mathbf{u} \in \mathcal{G}_{W^c}.
\label{compact}
\end{equation}

As mentioned before, problem (\ref{compact}) is a joint scheduling and bandwidth provisioning problem. A possible solution approach is to first find a feasible scheduler that requires the least bandwidth possible, and then solve the resulting provisioning problem only for this scheduler. This observation leads us to the following proposition.

\begin{proposition}
Consider the optimization problem:
\begin{equation}
\underset{W^c, \mathbf{u}}{\text{minimize}} \: \: W^c \: \: s.t. \: \: \mathbf{u} \in \mathcal{F}_{W^c}.
\label{compact-g}
\end{equation}
Suppose $\exists \{\mathbf{u}_{W^c}\}_{W^c \in \mathbb{R}}$ s.t. $\forall W^c \in \mathbb{R}$, if $\mathbf{u}_{W^c} \notin \mathcal{F}_{W^c}$, then $\mathcal{F}_{W^c}=\emptyset$. Next, consider optimization problem:
\begin{equation}
\underset{W^c}{\text{minimize}} \: \: W^c \: \: s.t. \: \: \mathbf{u}_{W^c} \in \mathcal{F}_{W^c}.
\label{schedprob3}
\end{equation}
If $W^{c*}$ optimally solves (\ref{schedprob3}), then $(W^{c*}, \mathbf{u}_{W^{c*}})$ optimally solves (\ref{compact-g}). Also, if $(W^{c*}, \mathbf{v}^*)$ optimally solves (\ref{compact-g}), then $W^{c*}$ optimally solves (\ref{schedprob3}).

\noindent Proof: See Appendix \ref{appc}.
\label{equiv2}
\end{proposition}

Proposition \ref{equiv2} implies that if there exists a set of schedulers parameterized by $W^c$ s.t. for every bandwidth $W^c$, either its scheduler corresponding to $W^c$ can satisfy the constraints or no other scheduler can, then it suffices to consider only that set of schedulers and solve only for $W^c$ as shown in (\ref{schedprob3}).

Thus, our goal now is to identify a scheduler that satisfies the premise of Proposition \ref{equiv2} for $\mathcal{F}_{W^c}=\mathcal{G}_{W^c}$, and then solve provisioning problem (\ref{schedprob3}) for that scheduler for $\mathcal{F}_{W^c}=\mathcal{G}_{W^c}$. Therefore, the original joint scheduling and bandwidth provisioning problem can be divided into a scheduling problem and a bandwidth provisioning problem. 

In Sec. \ref{schedulingprob}, we show that the well-known Max-Weight scheduler \cite{neely} solves the scheduling problem, i.e., it satisfies the premise of Proposition \ref{equiv2} for $\mathcal{F}_{W^c}=\mathcal{G}_{W^c}$, under the assumption that bandwidth demands $\mathbf{W}(t)$ follow an ergodic MC. Although in practice the markovian assumption may be violated, it still motivates us to use the Max-Weight scheduler in general settings. Later in Sec. \ref{dimensioningprob}, we address the bandwidth provisioning problem for the Max-Weight Scheduler.

\section{Scheduling Problem}
\label{schedulingprob}
To identify a set of schedulers that satisfies the premise of Proposition \ref{equiv2}, note that the second and third constraint in (\ref{schedprob2}) specify a set of allowed control actions $\mathcal{U}(t)$ that depends on $\mathbf{W^e}(t)$ and $W^c$. Thus, problem (\ref{schedprob2}) can be rewritten as:
\begin{align}
& \underset{W^c,\mathbf{u}}{\rm{minimize}} \qquad W^c \nonumber\\
& \text{s.t.:} \: \liminf\limits_{T \to \infty}\frac{1}{T}\sum\limits_{t=1}^{T}u_i(t)\mathbbm{1}_{W_i^e(t) >0} \geq P_i^H - P_i^M,  \: w.p.1, \: \forall i, \nonumber\\
& \phantom{\text{s.t.:} \:} \mathbf{u}(t) \in \mathcal{U}(\mathbf{W^e}(t),W^c), \forall t.
\label{schedprob4}
\end{align}

Given that schedulers operate online, we consider only causal schedulers, i.e., schedulers that at time $t$ do not know the future bandwidth demands $W_i(\tau)$, $\tau > t$. Therefore at time $t$, the control decisions $\mathbf{u}(t)$ may depend only on past stored information such as $W^e_i(\tau)$ and $\mathbf{u}(\tau)$ where $\tau < t$. Here, we consider that the scheduler computes $\mathbf{u}(t)$ based on a state $\mathbf{x}(t)$ that summarizes the system's history at time $t$.

To determine the information that the state $\mathbf{x}(t)$ should include, note that the knowledge of $\mathbf{W^e}(t)$ and $W^c$ is needed, otherwise the set of feasible control actions cannot be found. Thus, $(\mathbf{W^e}(t),W^c) \in \mathbf{x}(t)$. Next, note that the first constraint (\ref{schedprob4}) is a time-average constraint. Thus, Lyapunov optimization is applicable \cite{neely}. For this reason, we introduce the following:
\begin{equation}
d_i(t+1) = [ d_i(t) -u_i(t)\mathbbm{1}_{W_i^e(t)>0}]^+ + P_i^H-P_i^M, \: \forall i.
\label{deficits}
\end{equation}

Quantities $d_i(t)$ can be viewed as deficits; in each timeslot the deficit to NS $i$ increases by $P^H_i-P^M_i$ unless there is a positive excess demand and the scheduler satisfies it, i.e.,  $W_i^e(t)>0$ and $u_i(t)=1$, in which case the deficit decreases. Deficits have been widely used for scheduling packets with deadlines where the goal is to meet delivery ratios \cite{ghaderi}.

Deficits $d_i(t)$ can also be viewed as virtual queue lengths where in each timeslot, the number of arrivals is $P_i^H - P_i^M$ and the amount of service received is $u_i(t)\mathbbm{1}_{W_i^e(t)>0}$ for each queue. The introduction of deficits $d_i(t)$ is useful due to the following implication \cite[Theorem 2.8b]{neely}:
\begin{equation}
\begin{aligned}
& \limsup\limits_{T \to \infty}\sum\limits_{t=1}^T\E[d_i(t)] < \infty\\
\Rightarrow & \lim\limits_{T \to \infty}\frac{1}{T}\sum\limits_{t=1}^{T}u_i(t)\mathbbm{1}_{W_i^e(t) > 0} \geq P^H_i - P^M_i,\: w.p.1.
\end{aligned}
\label{rate-stability}
\end{equation}

Implication (\ref{rate-stability}) states that if deficits $d_i(t)$ are strongly stable, then the time-average constraints in (\ref{schedprob4}) are satisfied. Consequently, the knowledge of deficits $\mathbf{d}(t)$ may be useful to the scheduler. Thus, we consider schedulers that produce $\mathbf{u}(t)$ at each timeslot $t$ based on $\mathbf{x}(t)=(\mathbf{W^e}(t),\mathbf{d}(t),W^c)$.

Now, consider a scheduler that observes $\mathbf{x}(t)$ and chooses to satisfy the combination of NSs with the largest sum of deficits given the available bandwidth.
Equivalently, consider a scheduler whose decisions $\mathbf{u}^*_{W^c}(t)$ are obtained by solving:
\begin{align}
& \underset{\mathbf{u}(t)}{\text{maximize}} \qquad \sum\limits_{i: W_i^e(t) >0}d_i(t)u_i(t) \nonumber\\
& \text{s.t.:} \: \sum\limits_{i: W_i^e(t) >0}u_i(t)W_i^e(t) \leq W^c + \sum\limits_{i: W_i^e(t) <0}|W_i^e(t)|, \nonumber\\
& \phantom{\text{s.t.:} \:} \mathbf{u}(t) \in \{0,1\}^N.
\label{max-weight}
\end{align}

The scheduler obtained by solving (\ref{max-weight}) at each timeslot $t$ is called the Max-Weight scheduler. Note that (\ref{max-weight}) is a binary knapsack problem in each timeslot $t$. Although, the binary knapsack problem is known to be NP-Hard, there exist fully polynomial time approximation schemes. Thus, in practice, close to optimal solutions can be obtained fast.

Intuitively, we expect that if the Max-Weight scheduler $\mathbf{u}^*_{W^c}$ cannot stabilize the deficits for bandwidth $W^c$, then no other scheduler can stabilize them for bandwidth $W^c$. Indeed, it can be shown that the Max-Weight scheduler satisfies the premise in Proposition \ref{equiv2} when the bandwidth demands $\mathbf{W}(t)$ follow an ergodic MC \cite[Chapter 4.9.2]{neely}. Thus, it holds:
\begin{proposition}
Let $\mathbf{u}^*_{W^c}$ be the Max-Weight scheduler for bandwidth $W^c$ obtained by solving (\ref{max-weight}) for $\forall t$. Suppose $\mathbf{W}(t)$ follows an ergodic MC. If $\mathbf{u}^*_{W^c} \notin \mathcal{G}_{W^c}$, then $\mathcal{G}_{W^c} = \emptyset$.

\noindent Proof: See Appendix \ref{bigproof}.
\label{main-body-mw}
\end{proposition}

We note that the Max-Weight scheduler not only satisfies the premise of Proposition \ref{equiv2} for $\mathcal{F}_{W^c}=\mathcal{G}_{W^c}$ as stated in Proposition \ref{main-body-mw}, but also it does not require any knowledge of the statistics of bandwidth demands $\mathbf{W}(t)$. This is a major advantage since in case of incorrect estimation of the statistics of $\mathbf{W}(t)$, the scheduler does not need to be re-adjusted.

\section{Bandwidth Provisioning Problem}
\label{dimensioningprob}
Due to Proposition \ref{equiv2} and Proposition \ref{main-body-mw}, the original joint scheduling and bandwidth provisioning problem is equivalent to a bandwidth provisioning problem for the Max-Weight scheduler, if the bandwidth demands $\mathbf{W}(t)$ are markovian. Consequently, in that case, it suffices to solve:
\begin{equation}
\begin{aligned}
& \underset{W^c}{\rm{minimize}} \qquad W^c  \qquad \text{s.t. } \forall i \text{:}\\
& \liminf\limits_{T \to \infty}\frac{1}{T}\sum\limits_{t=1}^{T}u^{*}_{W^c,i}(t)\mathbbm{1}_{W_i^e(t) >0} \geq P_i^H \! - \! P_i^M,  \: w.p.1.\\
\end{aligned}
\label{dimprob}
\end{equation}

Although we proved the optimality of the Max-Weight scheduler under the assumption of ergodic markovian demands $\mathbf{W}(t)$, it still motivates us to consider the Max-Weight scheduler in general settings, where the markovian assumption may not hold. Thus, in all cases, we are interested in solving (\ref{dimprob}).

However, the derivative of the probability in the constraint w.r.t. $W^c$ is not readily available. Thus, derivative-free methods are needed. Since $u^*_{W^c,i}(t)$ is increasing w.r.t. $W^c$, then (\ref{dimprob}) is a monotonic optimization problem. Thus, binary search or polyblock outer approximation \cite{monotonic} can be used to solve it.

\section{Overall Solution Approach}
\label{sola}
Here, we summarize the overall solution approach for the original joint scheduling and bandwidth provisioning problem (\ref{schedprob1}). Regarding scheduling, we use the Max-Weight scheduler motivated by its optimality under ergodic markovian demands $\mathbf{W}(t)$. The scheduler's decisions are found by solving (\ref{max-weight}) online at each timeslot $t$. Since (\ref{max-weight}) is a binary knapsack problem, we can solve it fast using approximation algorithms.

Next, we need to find quantities $W_i^L$ and $P_i^M$ to solve provisioning problem (\ref{dimprob}). To do so, we consider that whenever a new NS needs to be deployed, a data collection period begins during which we observe previous bandwidth demands and generate future bandwidth demands $\{\mathbf{W}(t)\}_{t \in \mathbb{N}}$ based on some traffic forecasting model. Then, since $F_{W_i}$ is an increasing function, we find $W_i^L$ by applying binary search on this sequence $\{\mathbf{W}(t)\}_{t \in \mathbb{N}}$. Note that during the process of finding $W_i^L$, we also obtain $P^M_i$ as follows from (\ref{wl}).

Once $W_i^L$ and $P^M_i$ are obtained, problem (\ref{dimprob}) can be solved using binary search. Once this is done, both the scheduler $\mathbf{u}$ and the required bandwidths $\mathbf{W^r}=\mathbf{W^L}$, $W^c$ are specified. Thus, a feasible solution to (\ref{schedprob1}) has been found. In case demands the sequence $\{\mathbf{W}(t)\}_{t \in \mathbb{N}}$ was generated by an ergodic MC, this feasible solution is the optimal solution. 

This overall process is repeated each billing cycle to update the forecasting model and charge each tenant accordingly. Some forecasting models for RANs can be found in \cite{forecasting1,forecasting2}.

\section{Experimentation in ns-3}
\label{simrel}
\begin{figure}
\centering
\includegraphics[width=\linewidth]{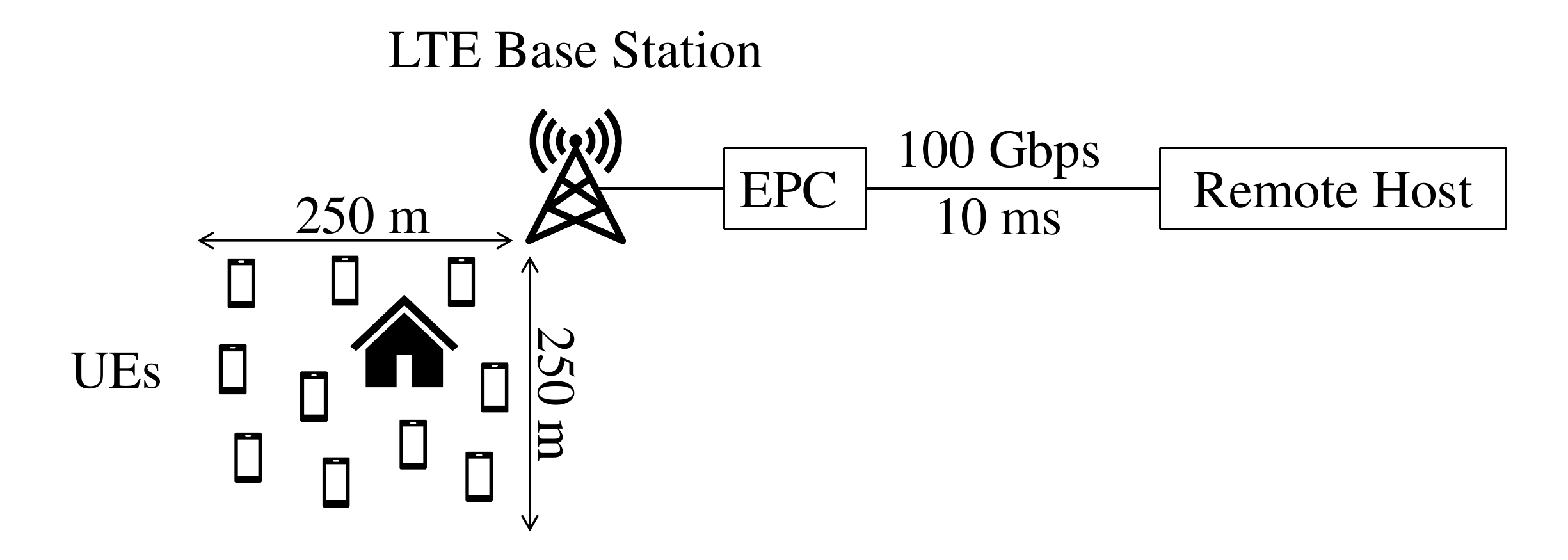}
\vspace{-0.8 cm}
\caption{In our ns-3 simulation setup, we consider a sector of an LTE BS covering a square area. A building was placed to simulate channel conditions in urban areas. The UEs move with 1.5m/s speed within this square and their propagation losses are computed by the "OhBuildings" model \cite[p. 40]{ns3}. The system bandwidth is 100 PRBs. We collect the bandwidth demands for each NS by executing an instance of this setup with a different number of UEs and a different application on the remote host generating downlink traffic.}
\label{netarc}
\end{figure}

\begin{figure}
\centering
\includegraphics[width=\linewidth]{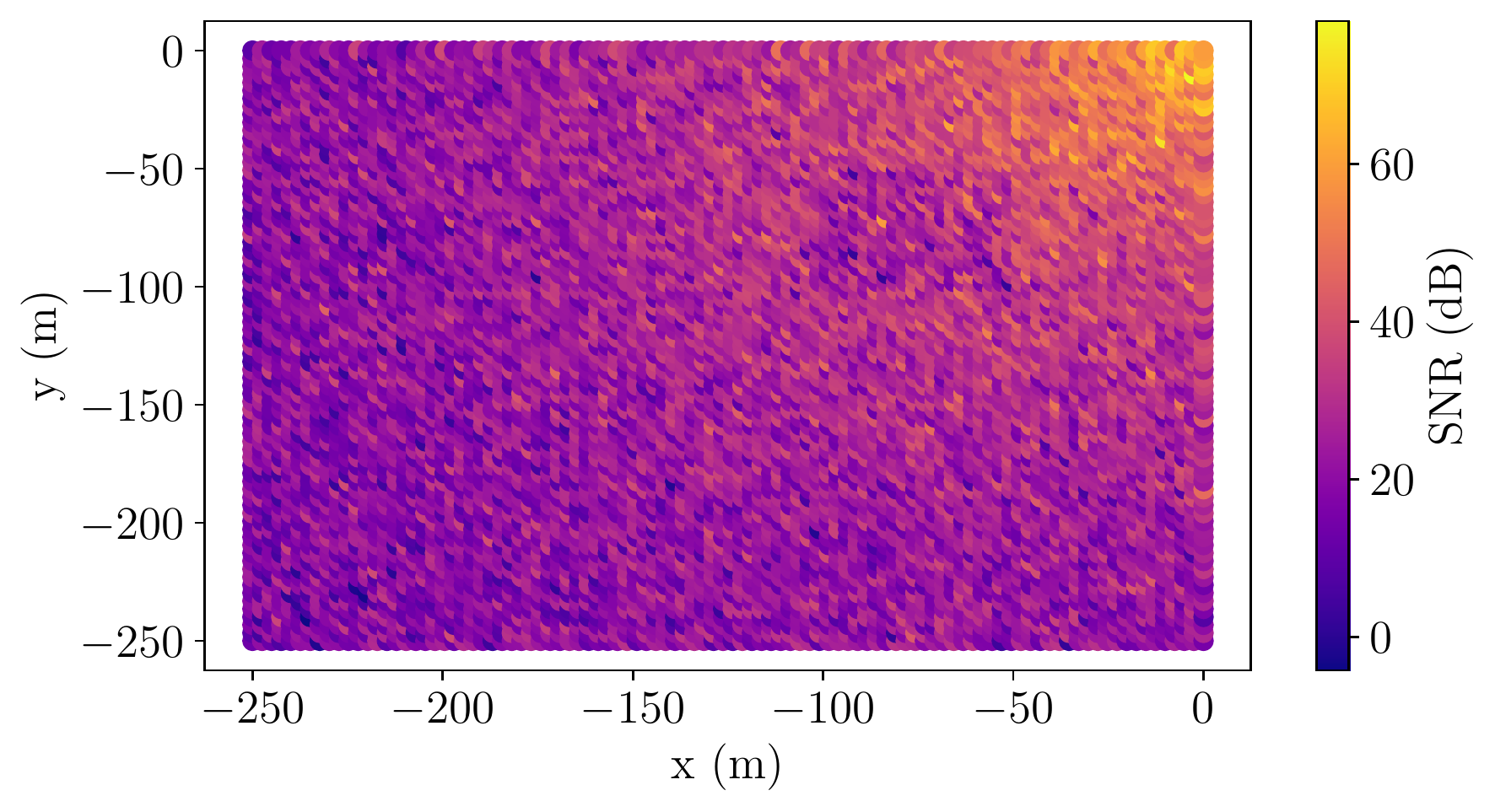}
\vspace{-0.8 cm}
\caption{The resulting REM of the considered network topology at 1m above the ground. The UEs move within this area and their downlink SNR changes accordingly. The BS is placed at (0, 0) and 20m above the ground. The building's dimensions are 50m x 50m x 10m and placed at (-100, -100).}
\label{REM}
\end{figure}

Here, we test our overall solution approach on bandwidth demands obtained using the LTE module of ns-3 \cite[Chapter 19]{ns3}. All simulations were run on a home computer with an Intel i7-10700K processor using 16 GB of RAM running on Windows 10. Our simulation setup is depicted in Fig. \ref{netarc} and the associated Radio Enviroment Map (REM) in Fig. \ref{REM}. We deploy one voice, one video streaming and two web-browsing NSs. The simulation time in all scenarios is 10 minutes due to time complexity. The User Equipment (UE) activity changes every few seconds instead of minutes to observe various traffic patterns within this short simulation time.

For the voice NS, we create 100 UEs receiving traffic from the remote host with 1kBps data rate and 20B packet size to simulate the G.729A codec. Each UE's call duration is modeled by a Pareto random variable with a mean of 2s. The idle duration of each UE follows the exponential distribution with a mean of 5s. Hence, each UE is active $29\%$ of the time. We model the traffic of the video streaming NS as previously, however we consider 15 UEs with average idle duration of 10s, 8Mbps data rate and 1kB packet length. 

The two web-browsing NSs were simulated by the default model in the 3GPP HTTP applications module of ns-3 \cite[Chapter 5]{ns3}. This model considers random byte sizes required to load objects of a website, random parsing times of these objects, and random reading times during which the UE remains idle. For the first web-browsing NS, we consider 10 UEs and for the second one 20 UEs.

Since we have not yet developed a bandwidth demand estimator as in Fig. \ref{system}, we approximate the bandwidth demand of a NS by the number of PRBs that the MAC scheduler allocated to the NS. Here, we encounter a challenge. Most MAC schedulers in ns-3 always allocate all 100 PRBs since it is wasteful not to do so \cite[Chapter 19.1.7]{ns3}. For instance, even if there is only one UE in the voice NS that requires just 1kBps to receive good QoS, all PRBs are allocated to that UE. Then, the bandwidth demand of each NS is a horizontal line at 100 PRBs which nullifies the benefits of multiplexing.

To tackle this issue, we need a MAC scheduler that does not always use all 100 PRBs. We found that the Token Bank Fair Queueing (TBFQ) MAC scheduler possesses this desired behavior \cite[p. 217]{ns3}. Specifically, this MAC scheduler stops allocating extra PRBs if the already allocated ones are enough to empty the packet buffers of all UEs. Thus, using the TBFQ MAC scheduler, we can approximate the resource demand of a NS when the desired QoS is to transmit all packets of all UEs within 1 ms. Although this is an extreme QoS requirement, it allows us to obtain bandwidth demands that vary over time.

\begin{figure}
\centering
\includegraphics[width=\linewidth]{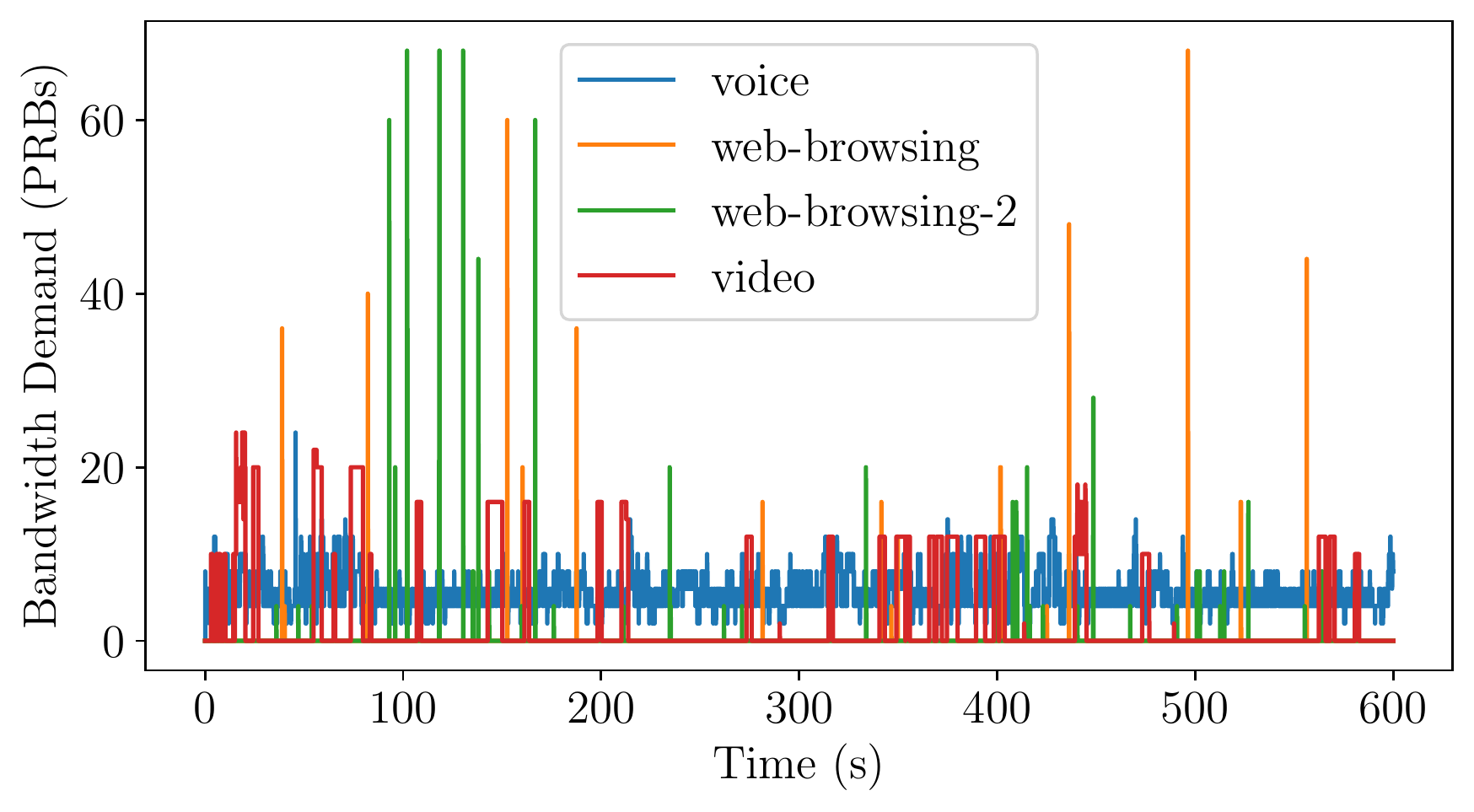}
\vspace{-0.8 cm}
\caption{The bandwidth demands of all four NSs for a 10 minute simulation. Since their peaks are not synchronized, multiplexing can be highly beneficial.}
\label{demands_100}
\end{figure}

Since the simulation time is 10 minutes, we obtain 600K bandwidth demands for each NS. Then, we find the maximum every 100 ms and obtain the bandwidth demands in Fig. \ref{demands_100}.

\begin{figure}
\centering
\includegraphics[width=\linewidth]{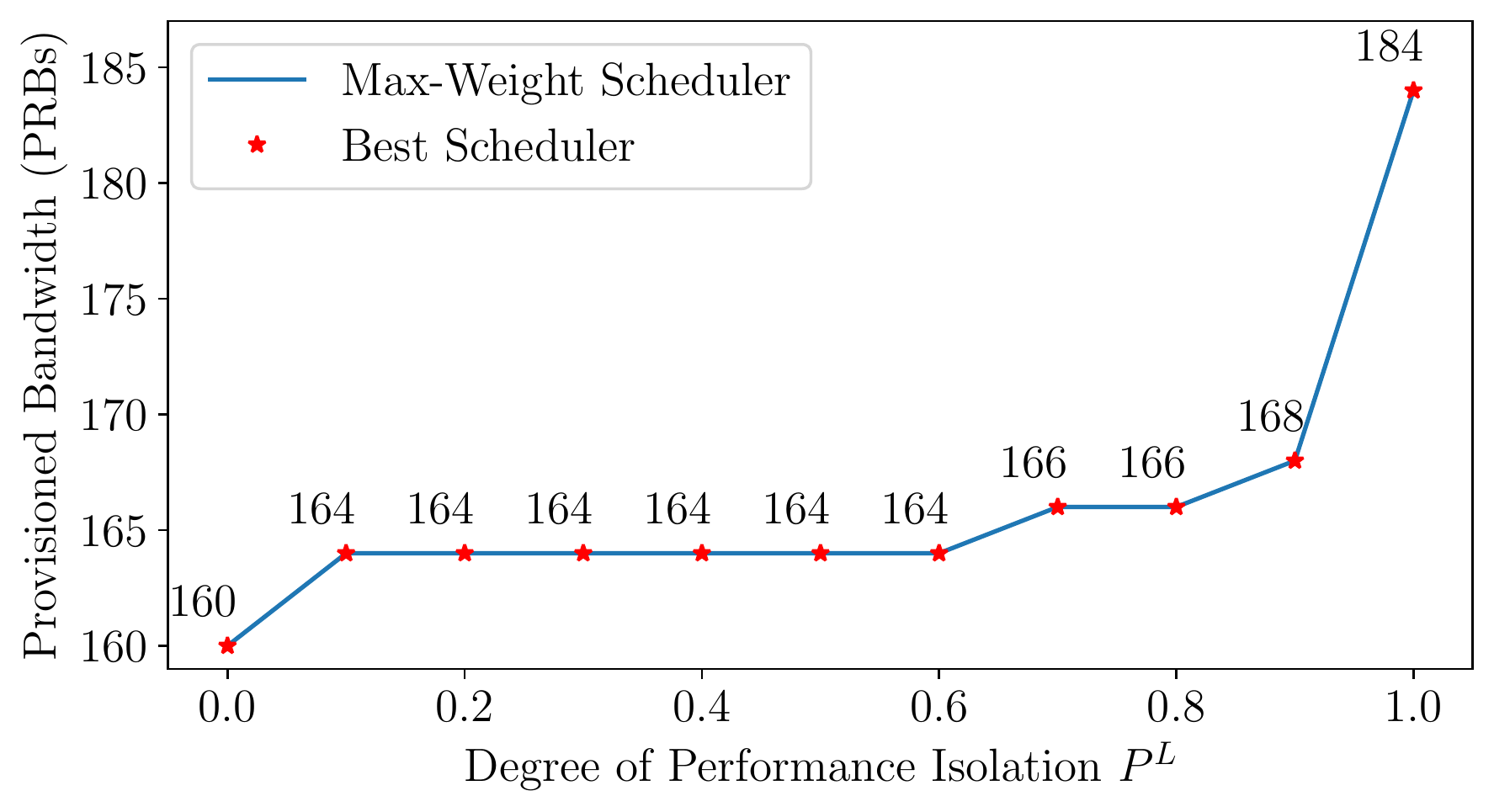}
\vspace{-0.8 cm}
\caption{Here, we consider $P^L=P^H=1$ for all NSs except from the voice NS whose degree of isolation $P^L$ we vary. The figure also shows that the Max-Weight scheduler requires as much bandwidth as the best scheduler, even though the bandwidth demands are not generated by a MC.}
\label{optimality_gap}
\end{figure}

\begin{figure}
\centering
\includegraphics[width=\linewidth]{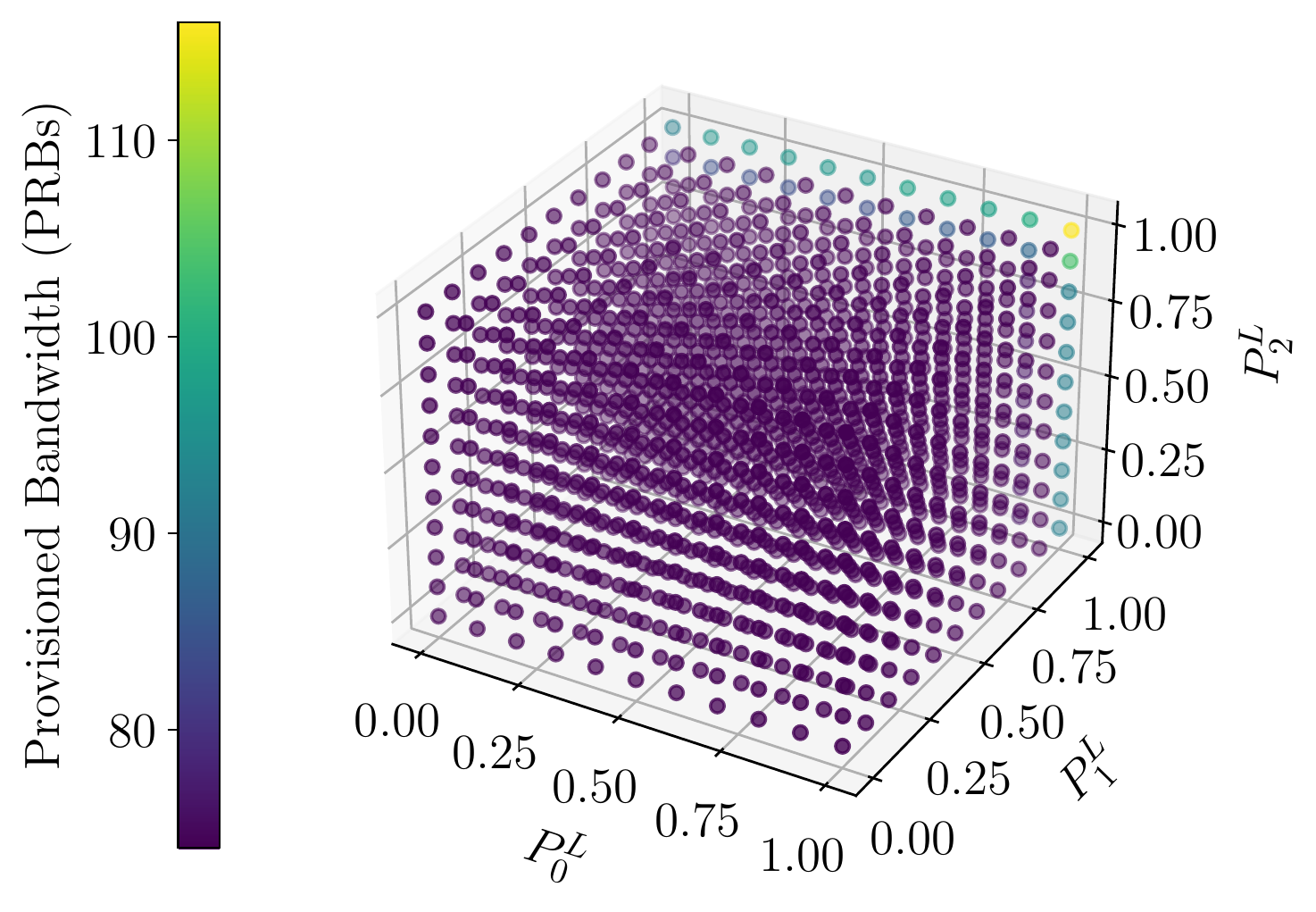}
\vspace{-0.8 cm}
\caption{The effect of the performance isolation vector $\mathbf{P^L}$ on the provisioned bandwidth. Point (0, 0, 0) corresponds to the case where no isolation is required and all NSs fully rely on multiplexing. Point (1, 1, 1) corresponds to full isolation for each NS where multiplexing is not allowed. The first case requires 74 PRBs and the second case requires 116 PRBs. Hence, multiplexing in this scenario reduces the provisioned bandwidth by $7.56$ MHz or by $36.2\%$.}
\label{simplex_tradeoff}
\end{figure}

Given the obtained sequence of bandwidth demands, we can study the tradeoff of interest by following Sec. \ref{sola} for various $P^H_i$ and $P^L_i$ values. Some such results are depicted in Fig. \ref{optimality_gap}. More extensive results are provided in Fig. \ref{simplex_tradeoff} where we consider the voice, web-browsing and video NSs.

In the Fig. \ref{optimality_gap}, we also compare the Max-Weight scheduler to the best scheduler. The best scheduler is found by solving (\ref{schedprob2}) for the obtained sequence of bandwidth demands which is a mixed integer linear program. To solve it, we use the scipy.optimize.milp function in python which uses the HiGHS optimization software \cite{highs}. Note that the best scheduler depends on the perfect forecasting of all future bandwidth demands and thus is non-causal. Clearly, its computation and performance is susceptible to errors in forecasting.

\begin{figure}
\centering
\includegraphics[width=\linewidth]{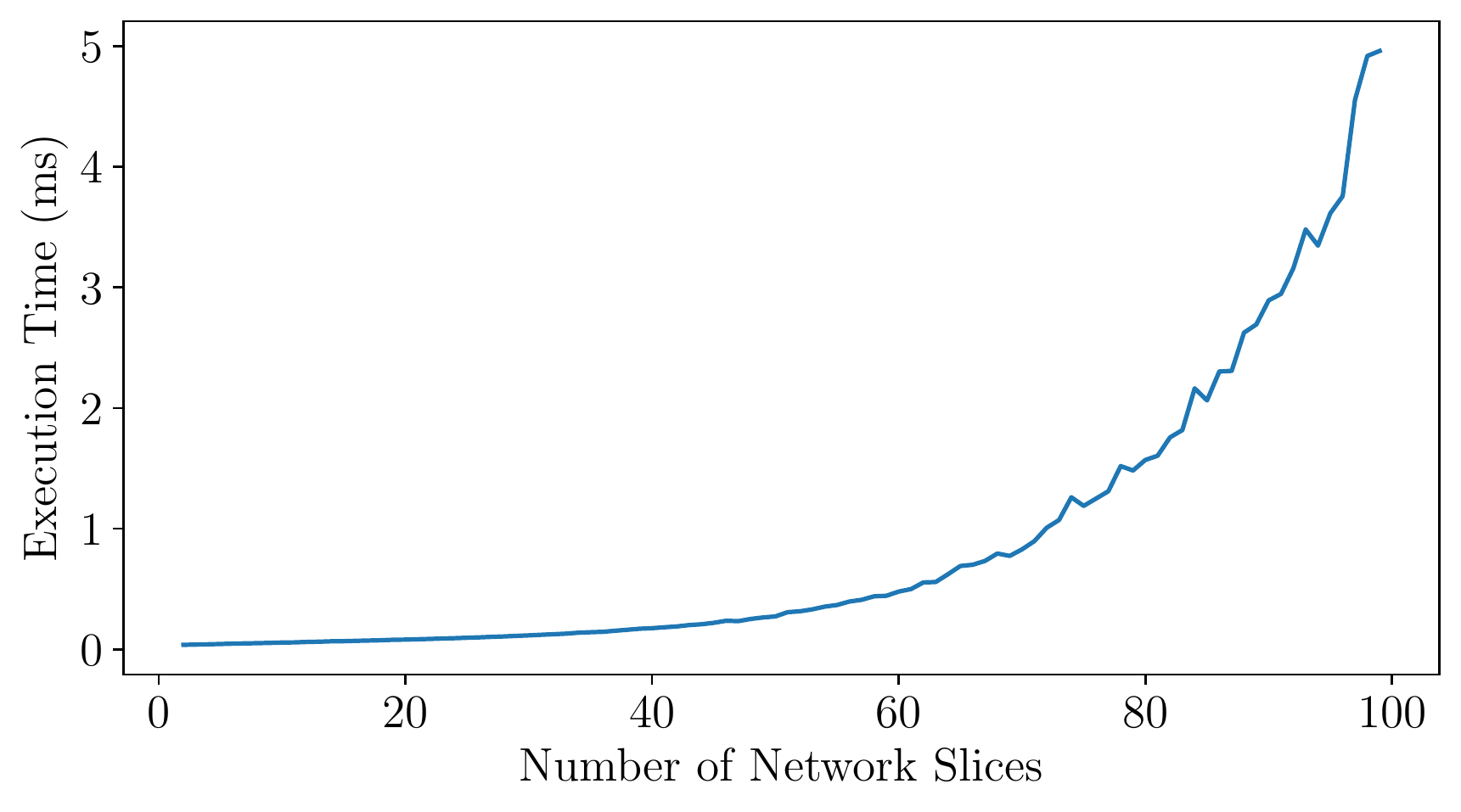}
\vspace{-0.8 cm}
\caption{The Max-Weight scheduler supports 1 ms time-granularity for a large number of NSs deployed at the same BS. Thus, the timeslot length in Fig. \ref{system} is affected by the timescale that the bandwidth demand estimator can support.}
\label{time_complexity}
\end{figure}

We finish this section by measuring the execution time of the Max-Weight scheduler, which requires the solution of a binary knapsack problem. In our implementation, we use the branch and bound algorithm from google OR-tools that obtains an optimal solution \cite{ortools}. The average execution time of the Max-Weight scheduler over 10K tests for a varying number of NSs and arbitrary bandwidth demands is shown in Fig \ref{time_complexity}.

\section{Conclusion}
\label{concl}
In this paper we studied the tradeoff between resource efficiency and performance isolation in network slicing. We showed that for bandwidth demands following an ergodic MC, the optimal multiplexing policy is given by the Max-Weight scheduler. We then tested the performance of the Max-Weight scheduler in some non-markovian settings and observed that it is still optimal. We also observed that the Max-Weight scheduler reduced the provisioned bandwidth by $36.2\%$ when each NS can fully rely on multiplexing.

We note that multiplexing in general requires the operation of a bandwidth demand estimator as in Fig. \ref{system}. The considerable bandwidth savings for the simple scenarios studied in this paper motivates the development of this network function. For future work, we wish to develop such a bandwidth demand estimator using online learning and data-driven methods.

\appendices
\section{Proof of Proposition \ref{simplify}}
\label{appa}
First, note that the first point $(W^c, \mathbf{W^L} +\mathbf{e},\mathbf{u})$ and the second point $(W^c + \mathbf{1}^{\top}\mathbf{e},\mathbf{W^L}, \mathbf{v})$ have the same objective function value. Thus, it suffices to show that if the first point satisfies the constraints, then the second point satisfies them as well. Clearly, if the first point satisfies the second and fourth constraints of (\ref{schedprob2}), then so does the second point. 

Next, we consider the third constraint. Suppose the first point satisfies it. We need to show that $(W^c + \mathbf{1}^{\top}\mathbf{e},\mathbf{W^L}, \mathbf{v})$ also satisfies it. Equivalently, by re-arranging terms and using the definition of $\mathbf{v}$, we need to show:
\begin{equation}
\begin{aligned}
& \sum_{i: u_i(t)=1,W_i(t) \geq W_i^L + e_i} W_i(t)-W_i^L \\
& \leq W^c + \sum_{i}e_i+\sum\limits_{i:W_i(t) \leq W^L_i + e_i}W^L_i-W_i(t).
\end{aligned}
\label{target}
\end{equation}
Note however that if the first point $(W^c, \mathbf{W^L} +\mathbf{e},\mathbf{u})$ satisfies the third constraint, it holds:
\begin{equation}
\begin{aligned}
& \sum\limits_{i: u_i(t)=1,W_i(t) \geq W_i^L + e_i}W_i(t)-W_i^L - e_i \\ 
& \leq W^c +\sum\limits_{i:W_i(t) < W^L_i + e_i}W^L_i + e_i -W_i(t).
\end{aligned}
\label{premise}
\end{equation}
We prove (\ref{target}) by moving the sum of $e_i$ in the left-side of (\ref{premise}) and by increasing its right-side by $\sum\limits_{i:u_i(t)=0,W_i(t) \geq W_i^L + e_i}e_i$.
Lastly, we need to show that if the first point satisfies the first constraint, so does the second one. From (\ref{assumption}), we need to show:
\begin{equation}
\begin{aligned}
\liminf\limits_{T \to \infty} & \frac{1}{T}\sum\limits_{t=1}^{T}\mathbbm{1}_{W_i(t) \leq W_i^L} + v_i(t)\mathbbm{1}_{W_i(t) > W_i^L}\\ 
\geq \liminf\limits_{T \to \infty} & \frac{1}{T}\sum\limits_{t=1}^{T}\mathbbm{1}_{W_i(t) \leq W_i^L + e_i}+u_i(t)\mathbbm{1}_{W_i(t) > W_i^L + e_i}.
\end{aligned}
\label{suffcond}
\end{equation}
By definition of $\mathbf{v}$, inequality (\ref{suffcond}) holds with equality.

\section{Proof of Proposition \ref{equiv1}}
\label{appb}
We start with the first statement. Since $(W^c,\mathbf{W}^r,\mathbf{u})$ feasible, then $\mathbf{W}^r=\mathbf{W^L} + \mathbf{e}$, where $\mathbf{e} \in \mathcal{R}^N_+$. Due to Proposition \ref{simplify}, it follows that $(W^c + \mathbf{1}^{\top}\mathbf{e},\mathbf{W^L}, \mathbf{v})$ is also an optimal solution to (\ref{schedprob1}). Next, note that the feasibility region of (\ref{schedprob2}) is a subset of the feasibility region of (\ref{schedprob1}). Thus, since $(W^c + \mathbf{1}^{\top}\mathbf{e},\mathbf{W^L}, \mathbf{v})$ optimally solves (\ref{schedprob1}), then $(W^c + \mathbf{1}^{\top}\mathbf{e}, \mathbf{v})$ must be an optimal solution to (\ref{schedprob2}).

We prove the second statement by contradiction. Suppose $\exists (W^{c'},\mathbf{W^{r'}},\mathbf{u'})$ s.t. $W^{c'} + \mathbf{1}^\top \mathbf{W^{r'}} < W^c + \mathbf{1}^\top \mathbf{W^L}$. If $\mathbf{W^{r'}}=\mathbf{W^{L}}$, it follows that $(W^{c'},\mathbf{u'})$ is a feasible solution to (\ref{schedprob2}) with $W^{c'} < W^c$. Thus, $(W^c,\mathbf{u})$ does not optimally solve (\ref{schedprob2}), which is a contradiction. If $\mathbf{W^{r'}}=\mathbf{e}+\mathbf{W^{L}}$, where $\mathbf{e} \in \mathcal{R}^N_+$, due to Proposition \ref{simplify}, $\exists (W^{c'} + \mathbf{1}^{\top}\mathbf{e},\mathbf{W^L}, \mathbf{v})$ that optimally solves (\ref{schedprob1}). Thus, $(W^{c'}+\mathbf{1}^{\top}\mathbf{e}, \mathbf{v})$ feasible solution to (\ref{schedprob2}) with $W^{c'}+\mathbf{1}^{\top}\mathbf{e} \leq W^c$, which is a contradiction.

\section{Proof of Proposition \ref{equiv2}}
\label{appc}
Suppose the first statement does not hold, i.e., $\exists (W^{c'},\mathbf{u}')$ that optimally solves (\ref{compact-g}), where $W^{c'} < W^{c*}$. Then, $\mathbf{u'} \in \mathcal{F}_{W^{c'}}$, thus $\Rightarrow \mathcal{F}_{W^{c'}} \neq \emptyset$. Due to the premise, it follows that $u_{W^{c'}} \in \mathcal{F}_{W^{c'}}$. Thus, $W^{c'}$ and $W^{c*}$ optimally solve (\ref{schedprob3}) which is a contradiction. For the second statement, $(W^{c*}, \mathbf{v}^*)$ optimally solves (\ref{compact-g}), thus $\mathcal{F}_{W^{c*}} \neq \emptyset$. Due to the premise, it follows that $\mathbf{u}_{W^{c*}} \in \mathcal{F}_{W^{c*}}$. Thus, $(W^{c*}, \mathbf{u}_{W^{c*}})$ also optimally solves (\ref{compact-g}). Therefore, $W^{c*}$ optimally solves (\ref{schedprob3}).

\input proof.tex

\bibliography{references}
\bibliographystyle{IEEEtran}

\end{document}

%% file: proof.tex
\section{Proof of Proposition \ref{main-body-mw}}
\label{bigproof}
Our proof follows the same procedure described over several chapters in \cite{neely} and is divided into two parts.

First, we consider a class of schedulers that make decisions based solely on the current bandwidth demands $\mathbf{W^e}(t)$ called $\mathbf{W^e}(t)$-only schedulers. Using \cite[Theorem 4.5]{neely}, we show that there exists a set of schedulers of this class that satisfies the premise of Proposition \ref{equiv2} for ergodic markovian demands.

Second, by following the steps in \cite[p.34-p.36]{neely}, we show that the Max-Weight scheduler strongly stabilizes the deficits $d_i(t)$ whenever that set of $\mathbf{W^e}(t)$-only schedulers stabilizes them. Thus, we show that the Max-Weight scheduler also satisfies the premise of Proposition \ref{equiv2} for ergodic markovian demands. 

The step-by-step details of the first part and the second part of this proof are in Appendix \ref{firstpart} and Appendix \ref{secondpart} respectively.

\subsection{$\mathbf{W^e(t)}$-only schedulers}
\label{firstpart}
Here, we show that there exists a set of $\mathbf{W^e}(t)$-only schedulers that satisfies the premise of Proposition \ref{equiv2} when $\mathbf{W}(t)$ follows an ergodic MC. Suppose $\mathbf{W}(t)$ follows an ergodic MC, and let $\pi_i^w$ denote the probability of state $w \in \mathcal{S}_i$, where $\mathcal{S}_i$ the state space of MC $i$. Since each MC $i$ is ergodic, it follows \cite[Corollary 9.29]{queuebook}:
\begin{equation}
 \liminf\limits_{T \to \infty}\frac{1}{T}\sum\limits_{t=1}^{T}\mathbbm{1}_{W_i(t) \leq W_i^r}=\sum_{w \in \mathcal{S}_i : w \leq W_i^r}\pi_i^w, \: w.p.1, \: \forall i.
\label{time-averages}
\end{equation}

From (\ref{time-averages}), it follows that assumption (\ref{assumption}) holds. Moreover, since the MC of $\mathbf{W}(t)$ is ergodic, its limiting distribution is the stationary distribution \cite[Theorem 9.4 and Theorem 8.6]{queuebook}. Therefore, regardless of the initial distribution of the MC, after some period of time, the process $\mathbf{W}(t)$ becomes identically distributed (id). 

Assuming that this transition period is small enough, we consider that the process once it has reached its stationary distribution. Thus, $\Pr(W_i(t)=w)=\pi_i^w, \,\forall t \in \mathbb{N}$, where $\pi_i$ is the stationary distribution of NS $i$ and $\pi_i^w$ is the probability of state $w$. Hence, the vector $\mathbf{W^e}(t)$ is also id with $\Pr(W_i^e(t)=w-W^L_i)=\pi_i^w, \, \forall w \in \mathcal{S}_i$. Let $\mathcal{S}_i^e \triangleq \{w^e: w^e+W^L_i \in \mathcal{S}_i\}$, $\mathcal{S}^e \triangleq \mathcal{S}_1^e \times ... \times \mathcal{S}_N^e$ and $p_i^{w^e} \triangleq \pi_i^{w^e+W^L_i}$, $\forall w^e \in \mathcal{S}_i^e$. Then, $\Pr(\mathbf{W^e}(t)=\mathbf{w^e})=\prod\limits_{i \leq N}p_i^{w^e_i} \triangleq p_{\mathbf{W^e}}(\mathbf{w^e})$, where $\mathbf{w^e} \in \mathcal{S}^e$.

Now, consider the class of $\mathbf{W^e}(t)$-only schedulers defined by probability distributions $p_{\mathbf{u}|\mathbf{w^e}}(\mathbf{u},\mathbf{w^e}) \triangleq \Pr(\mathbf{u}(t)=\mathbf{u}| \mathbf{W^e}(t)=\mathbf{w^e})$, i.e., if $\mathbf{W^e}(t)=\mathbf{w^e}$, then $\mathbf{u}(t)=\mathbf{u}$ with probability $p_{\mathbf{U}|\mathbf{W^e}}(\mathbf{u},\mathbf{w^e})$ which does not depend on time $t$. Note that $p_{\mathbf{U}|\mathbf{W^e}}(\mathbf{u},\mathbf{w^e})$ must satisfy:
\begin{equation}
\sum\limits_{\mathbf{u} \in \mathcal{U}(\mathbf{w^e},W^c)}p_{\mathbf{U}|\mathbf{W^e}}(\mathbf{u},\mathbf{w^e})=1, \: \forall \mathbf{w^e} \in \mathcal{S}^e.
\end{equation}
Next, we wish to compute the limit of the time average of $u_i(t)\mathbbm{1}_{W_i^e(t) > 0}$. To this end, first note that:
\begin{equation}
\begin{aligned}
& \E[u_i(t)\mathbbm{1}_{W_i^e(t) > 0}]= \bar{u_i} \triangleq\\
& \sum\limits_{\mathbf{w^e} \in \mathcal{S}^e, \mathbf{u} \in \mathcal{U}(\mathbf{w^e},W^c)}p_{\mathbf{U}|\mathbf{W^e}}(\mathbf{u},\mathbf{w^e})p_{\mathbf{W^e}}(\mathbf{w^e})u_i\mathbbm{1}_{W_i^e > 0}, \: \forall t.
\end{aligned}
\label{expectation}
\end{equation}
Since $u_i(t)\mathbbm{1}_{W_i^e(t) > 0}$ is modulated by the ergodic MC that $\{\mathbf{W^e}(t)\}_{t \in \mathbb{N}}$ follows, it holds \cite[p.76]{neely}:
\begin{equation}
\begin{aligned}
\liminf\limits_{T \to \infty}\frac{1}{T}\sum\limits_{t=1}^{T}u_i(t)\mathbbm{1}_{W_i^e(t) >0}=\frac{\E[\sum\limits_{t=1}^{T_1}u_i(t)\mathbbm{1}_{W_i^e(t) > 0}]}{\E[T_1]}, \: w.p.1,
\end{aligned}
\label{limit1}
\end{equation}
where $T_1$ is the first recurrence time to some arbitrary state $s \in \mathcal{S}^e$. Since $u_i(t)\mathbbm{1}_{W_i^e(t) > 0}$ is a bounded function of a finite-state and ergodic MC, the extension of Wald's equation for MCs in \cite{Moustakides} can be applied. Since we observe our MC after convergence to the stationary distribution, then the extension in \cite{Moustakides} results to the standard Wald's equation. Thus:
\begin{equation}
\liminf\limits_{T \to \infty}\frac{1}{T}\sum\limits_{t=1}^{T}u_i(t)\mathbbm{1}_{W_i^e(t) >0}=\bar{u_i}, \: w.p.1.
\label{limit2}
\end{equation}
Therefore, under an $\mathbf{W^e}(t)$-only scheduler, the long-term time-average of $u_i(t)\mathbbm{1}_{W_i^e(t) > 0}$ is equal to its one-slot expected value which is constant since $u_i(t)\mathbbm{1}_{W_i^e(t) > 0}$ is id. Now, consider the following constraints:
\begin{equation}
\begin{aligned}
& \liminf\limits_{T \to \infty}\frac{1}{T}\sum\limits_{t=1}^{T}\E[u_i(t)\mathbbm{1}_{W_i^e(t) >0}] \geq P_i^H - P_i^M,  \; \forall i,\\
& \mathbf{u}(t) \in \mathcal{U}(\mathbf{W^e}(t),W^c), \forall t.
\end{aligned}
\label{meanrate}
\end{equation}
Note that (\ref{meanrate}) is identical to the feasibility region of (\ref{schedprob4}) with the exception of the expected value operator $\E[\cdot]$ in the limit. Due to $(\ref{limit2})$, it can be shown that if there does not exist a $\mathbf{W^e}(t)$-only scheduler that satisfies (\ref{meanrate}), no scheduler satisfies (\ref{meanrate}) including non-causal schedulers \cite[Theorem 4.5]{neely}. 

Thus, Proposition \ref{equiv2} applies for a set of $\mathbf{W^e}(t)$-only schedulers, where $\mathcal{F}_{W^c}$ is defined by (\ref{meanrate}). Ideally, however we would like to show that it also applies for $\mathcal{F}_{W^c}=\mathcal{G}_{W^c}$. To this end, note that due to Fatou's lemma, it holds:
\begin{equation}
\begin{aligned}
& \liminf\limits_{T \to \infty}\frac{1}{T}\sum\limits_{t=1}^{T}u_i(t)\mathbbm{1}_{W_i^e(t) >0} \geq P_i^H - P_i^M, \: w.p.1 \\
\Rightarrow  & \liminf\limits_{T \to \infty}\frac{1}{T}\sum\limits_{t=1}^{T}\E[u_i(t)\mathbbm{1}_{W_i^e(t) >0}] \geq P_i^H - P_i^M, \: w.p.1.
\end{aligned}
\label{fatou}
\end{equation}
Using the above, we can show the following lemma.
\begin{lemma}
There exists $\mathcal{C'}\triangleq \{\mathbf{u}_{W^c}\}_{W^c \in \mathbb{R}}$ s.t. $\forall W^c \in \mathbb{R}$, if $\mathbf{u}_{W^c} \notin \mathcal{G}_{W^c}$, then $\mathcal{G}_{W^c}=\emptyset$, where each $\mathbf{u}_{W^c}$ is a $\mathbf{W^e}(t)$-only scheduler.

\noindent Proof:

We prove the contrapositive. Suppose $\exists \mathbf{v} \in \mathcal{G}_{W_c}$. Due to (\ref{fatou}), scheduler $\mathbf{v}$ satisfies (\ref{meanrate}). Thus, due to \cite[Theorem 4.5]{neely}, $\exists \mathbf{u}_{W^c}$ that satisfies (\ref{meanrate}) where $\mathbf{u}_{W^c}$ is a $\mathbf{W^e}(t)$-only scheduler. Thus, since (\ref{fatou}) holds with equivalence for $\mathbf{W^e}(t)$-only schedulers, $\mathbf{u}_{W^c} \in \mathcal{G}_{W_c}$ where $\mathbf{u}_{W^c}$ is $\mathbf{W^e}(t)$-only.

\label{adequacy}
\end{lemma}

Therefore, due to Proposition \ref{equiv2}, it is adequate to consider the set $\mathcal{C'}$ of $\mathbf{W^e}(t)$-only schedulers. Note that Lemma \ref{adequacy} shows the existence of set $\mathcal{C'}$ but it does not describe a method to find it. To find each $\mathbf{W^e(t)}$-only scheduler $\mathbf{u}_{W^c} \in \mathcal{C'}$, we need to solve the following problem for each $W^c \in \mathbb{R}$:
\begin{align}
& \underset{\epsilon_{W^c}, p_{\mathbf{U}|\mathbf{W^e}}}{\rm{maximize}} \qquad \epsilon_{W^c} \nonumber\\
& \text{s.t.:} \: \sum\limits_{\mathbf{u} \in \mathcal{U}(\mathbf{w^e},W^c)}p_{\mathbf{U}|\mathbf{W^e}}(\mathbf{u},\mathbf{w^e})=1, \: \forall \mathbf{w^e} \in \mathcal{S}^e, \nonumber \\
& \phantom{\text{s.t.:} \:} p_{\mathbf{U}|\mathbf{W^e}}(\mathbf{u},\mathbf{w^e}) \geq 0, \: \forall \mathbf{w^e} \in \mathcal{S}^e,\: \forall \mathbf{u} \in \mathcal{U}(\mathbf{w^e},W^c), \nonumber\\
& \phantom{\text{s.t.:} \:} \sum\limits_{\mathbf{w^e} \in \mathcal{S}^e, \mathbf{u} \in \mathcal{U}(\mathbf{w^e},W^c)}p_{\mathbf{U}|\mathbf{W^e}}(\mathbf{u},\mathbf{w^e})p_{\mathbf{W^e}}(\mathbf{w^e})u_i\mathbbm{1}_{W_i^e > 0} \nonumber \\
& \phantom{\text{s.t.:} \: \sum\limits_{\mathbf{w^e} \in \mathcal{S}^e, \mathbf{u} \in \mathcal{U}(\mathbf{w^e},W^c)}} \geq P_i^H - P_i^M + \epsilon_{W^c}, \: \forall i.
\label{findc}
\end{align}

In (\ref{findc}), the first two constraints ensure that the conditional distribution that defines the $\mathbf{W^e}(t)$-only scheduler is valid and the third constraint relates to the time-average constraint. Let $\epsilon^*_{W^c}$ denote the optimal value of (\ref{findc}). The optimal solutions of (\ref{findc}) for all $W^c \in \mathbb{R}$ compose set $\mathcal{C'}$. Clearly, if $\epsilon^*_{W^c}<0$, the long-term average constraint cannot be met by any $\mathbf{W^e}(t)$-only scheduler which means that no other scheduler can meet it as stated in Lemma \ref{adequacy}. Thus, if $\epsilon^*_{W^c}<0$, then $\mathcal{G}_{W^c} = \emptyset$.

\subsection{Max-Weight Scheduler}
\label{secondpart}
Our objective now is to show that the Max-Weight scheduler strongly stabilizes the deficits whenever $\epsilon^*_{W^c}<0$. If that is true, then it immediately follows from Lemma \ref{adequacy} and (\ref{rate-stability}) that the Max-Weight scheduler also satisfies the premise of Proposition \ref{equiv2} for ergodic markovian demands. To do so, we consider the following Lyapunov function:
\begin{equation}
L(\mathbf{d}(t)) \triangleq \frac{1}{2}\mathbf{1}^\top\mathbf{d}(t).
\label{lyapunov}
\end{equation}
Note that if $L(\mathbf{d}(t))$ is small then all deficits are small. Next, consider the conditional Lyapunov drift:
\begin{equation}
\Delta(\mathbf{d}(t)) \triangleq \E[L(\mathbf{d}(t+1))-L(\mathbf{d}(t))| \mathbf{d}(t)].
\label{drift}
\end{equation}
Note that if the conditional Lyapunov drift is small, then the deficits between two consecutive slots do not differ significantly. Following the same steps as in \cite[p.33]{neely}, we can show:
\begin{equation}
\begin{aligned}
\Delta(\mathbf{d}(t)) \leq  B + & \sum\limits_{i=1}^Nd_i(t)(P^H_i-P^M_i)\\
&-\E \left[\sum\limits_{i=1}^Nd_i(t)u_i(t)\mathbbm{1}_{W_i^e(t) > 0}|\mathbf{d}(t)\right],
\end{aligned}
\label{driftbound}
\end{equation}
where $B=1 + \sum_{i=1}^N(P^H_i-P^M_i)^2/N$. Note that since a small conditional drift implies "stable" deficits, we wish to reduce the conditional Lyapunov drift. To this end, we maximize the expectation in (\ref{driftbound}) by maximizing its argument. 

Thus, we need to derive a scheduler that observes the deficits $\mathbf{d}(t)$ and the bandwidth demands $\mathbf{W^e}(t)$, and maximizes the argument of the expectation in (\ref{driftbound}). This scheduler is precisely the Max-Weight scheduler defined in (\ref{max-weight}). Since the Max-Weight scheduler maximizes the expectation in (\ref{driftbound}), it holds:
\begin{equation}
\begin{aligned}
& \Delta(\mathbf{d}^*(t)) \leq B + \sum\limits_{i=1}^Nd_i^*(t)(P^H_i-P^M_i) \\
& -\E \left[\sum\limits_{i=1}^Nd^*_i(t)u_i(t)\mathbbm{1}_{W_i^e(t) > 0}|\mathbf{d}^*(t)\right], \: \forall \mathbf{u}(t) \in \mathcal{U}_{(\mathbf{w^e},W^c)},
\end{aligned}
\label{maxweightbound}
\end{equation}
where $\mathbf{d}^*(t)$ are the deficits under the Max-Weight scheduler. Since bound (\ref{maxweightbound}) holds for every possible scheduling decision $ \mathbf{u}(t) \in \mathcal{U}_{(\mathbf{w^e},W^c)}$, we can apply it for the $\mathbf{W^e}(t)$-only scheduler that solves (\ref{findc}). In that case, since $u_i(t)\mathbbm{1}_{W_i^e(t) > 0}$ is independent of deficits $\mathbf{d}^*(t)$ and  $\E[u_i(t)\mathbbm{1}_{W_i^e(t) > 0}]=P^H_i-P^M_i+\epsilon^*_{W^c}$, it holds:
\begin{equation}
\Delta(\mathbf{d}^*(t)) \leq B - \epsilon^*_{W^c}\sum\limits_{i=1}^Nd_i^*(t).
\label{newbound}
\end{equation}
By taking the expectation of (\ref{newbound}) and following the steps in \cite[p.36]{neely}, it holds:
\begin{equation}
\epsilon^*_{W^c}\limsup\limits_{T \to \infty}\frac{1}{T}\sum\limits_{t=1}^T\E[d^*_i(t)] \leq B, \: \forall i.
\label{almost-final}
\end{equation}

Therefore, if $\epsilon^*_{W^c}>0$, then all deficits $\mathbf{d}(t)$ are strongly stable. Due to (\ref{rate-stability}), it follows that the Max-Weight scheduler satisfies the time-average constraint whenever there exists a $\mathbf{W^e}(t)$-only scheduler that satisfies it. Let $\mathbf{u}'_{W^c}$ be the $\mathbf{W^e}(t)$-only scheduler defined by the solution of (\ref{findc}). Equivalently, if $\mathbf{u}'_{W^c} \in \mathcal{G}_{W^c}$, then $\mathbf{u}^*_{W^c} \in \mathcal{G}_{W^c}$. 

Given the above and Lemma \ref{adequacy}, it immediately follows that the Max-Weight scheduler satisfies the premise of Proposition \ref{equiv2} for $\mathcal{F}_{W^c}=\mathcal{G}_{W^c}$ when $\mathbf{W}(t)$ follows an ergodic MC. Therefore, our proof is finally complete.